# On the ambiguity of determination of interfering resonances parameters

V.M. Malyshev

*V.M.Malyshev@inp.nsk.su*

**Budker Institute of Nuclear Physics,
630090, Novosibirsk, Russia**

Abstract

The general form of solutions for parameters of interfering Breit-Wigner resonances is found. The number of solutions is determined by the properties of roots of corresponding characteristic equation and does not exceed $2^{N-1}$, where $N$ is the number of resonances. For resonances of more complicated form, provided that their amplitudes satisfy certain conditions, for any $N \geq 2$ multiple solutions also exist.

## 1. Introduction

In nuclear and high energy physics resonances are traditionally treated as simple poles of an $S$-matrix [1]. For example, the cross section of reaction
$$a + b \to R \to c + d,$$
which goes through an intermediate resonance $R$ of spin $J$, for unpolarized beams and in the absence of background is equal to
$$\sigma(E) = \frac{\pi}{p^2} \frac{2J+1}{(2S_1+1)(2S_2+1)} \left| a^{BW}(E) \right|^2,$$
where $E$ and $p$ are the c.m. energy and momenta of colliding particles of spins $S_1$ and $S_2$, $a^{BW}$ is a Breit-Wigner amplitude of the process. In a non-relativistic case this amplitude has the form
$$a^{BW}(E) = \frac{-\sqrt{\Gamma_i \Gamma_f}}{E - E_R + i\Gamma/2}, \tag{1}$$
where $\Gamma_i$ and $\Gamma_f$ are partial widths of resonance decays into initial and final states, respectively, $\Gamma$ is the total width of the resonance. Thus, the amplitude has a simple pole at $E = E_R - i\Gamma/2$ and the cross section is proportional (up to the factor $1/p^2$)
$$\sigma(E) \propto \left| a^{BW}(E) \right|^2 = \frac{\Gamma_i \Gamma_f}{(E - E_R)^2 + \Gamma^2/4} \propto BW(E), \tag{2}$$
where $BW(E)$ is the Breit-Wigner distribution,
$$BW(E) = \frac{1}{\pi} \frac{\Gamma/2}{(E - E_R)^2 + \Gamma^2/4}.$$
In a relativistic case the amplitude (1) is replaced by
$$a^{BW}(s) = \frac{-\sqrt{\Gamma_i \Gamma_f} \, m}{s - m^2 + im\Gamma}, \tag{3}$$
where $m$ is the mass of the resonance, $s$ is the c.m. energy squared.

Breit-Wigner amplitudes (1) and (3) well describe rather narrow resonances only. Broad resonances are usually asymmetric, so one has to use more complicated formulae for their description. For example, one can try to take into account the phase space of the final state using energy-dependent widths in amplitudes. Moreover, sometimes a reaction can go via several intermediate resonances and, as a rule, there is some non-resonant background. Therefore, in general case the cross section has the form

$$\sigma(E) \propto \left| \sum_j a_j(E) + b(E) \right|^2, \quad (4)$$

where $a_j(E)$ are the amplitudes of resonances, $b(E)$ is the amplitude of non-resonant background, i.e. resonances can interfere both between themselves and with the background.

In experiment, from the fit of data with the cross section (4) the resonance parameters, such as masses, widths, branching fractions and so on, are determined. If several resonances interfere (or there is interference with the background), apart from difficulties with the parameterization of asymmetric resonances, an additional problem arises. Since in experiment only the cross section of a reaction is measured, while the relative phases of resonances are usually unknown, the same energy dependence of the cross section can be obtained with different sets of parameters. Examples of such an ambiguity can be found, e.g., in Ref. [2], where double solutions for resonance parameters obtained in the analysis of experimental data for the processes $e^+e^- \to \pi^+\pi^- J/\psi$, $e^+e^- \to \pi^+\pi^- \psi(2S)$, $e^+e^- \to \omega\pi^0$, $e^+e^- \to \pi^+\pi^-$, are shown. In Refs. [3–5] the problem was investigated analytically. The authors of [3,4] have shown that for two interfering resonances as well as for a resonance interfering with a background, the number of solutions does not exceed two. A necessary and sufficient condition for the existence of a double solution was found. In Ref. [5] the problem was solved analytically for the case of two interfering Breit-Wigner resonances (BW-resonances). It was shown that there are two equivalent (i.e. giving the same energy dependence of the cross section) solutions and their explicit form was found. For three BW-resonances the problem was not solved analytically due to technical difficulties, however numerical experiments showed that in this case there are four equivalent solutions. It was suggested that in the case of $N$ resonances the number of solutions equals $2^{N-1}$. The situation changes for resonances with more complicated form of the amplitude. So, for a particular case of two interfering resonances with amplitudes of the form (3) and some kind of energy-dependent widths, it was analytically shown that there are no equivalent solutions. The numerical results of simulation for two and three such resonances were also presented. In fits of simulation data two and four solutions for the parameters, respectively, were found. However, these solutions are not equivalent – the solution, corresponding to the original set of parameters, gives the cross section which describes data better than additional solutions do.

In Ref. [5] the Fourier transform for determination of solutions was used. This led to cumbersome calculations and greatly complicated their search. At the same time the problem for an arbitrary number of Breit-Wigner resonances can be solved analytically in a more simple way. The number of equivalent solutions in general case is actually $2^{N-1}$, but it may be smaller. For resonances of a more complicated form, whose resonant behavior is due to pole singularities, it is found that for any $N \geq 2$ under certain conditions multiple solutions also exist.

## 2. The solutions for parameters of interfering resonances

Let us write the amplitude of the Breit-Wigner resonance (1) in the form

$$a^{\text{BW}}(E) = \frac{A}{E - M}, \quad (5)$$

where $A$ and $M$ are complex parameters. The amplitude is an analytic function in complex plane and has one simple pole, located near the interval $[E_1, E_2]$, where the cross section is measured. We shall write the resonance amplitudes of more complicated shapes similarly, explicitly writing out the resonant poles:

$$a(E) = \frac{Af(E)}{E - M}. \quad (6)$$

The function $f(E)$ will be assumed analytic in some region $\Omega$ containing the interval $[E_1, E_2]$. It may also have the poles which are located far enough from $[E_1, E_2]$.

With these notations the cross section of the reaction for $N$ interfering resonances takes the form:

$$\sigma(E) \propto \left|\sum_{k=1}^{N} a_k(E)\right|^2 = \left|\sum_{k=1}^{N} \frac{A_k f_k(E)}{E - M_k}\right|^2. \tag{7}$$

If there is a non-resonant background, its amplitude $Cf(E)$ ($C$ is a complex constant) should not obviously have poles near the interval $[E_1, E_2]$. The cross section in this case can be written in a similar form,

$$\sigma(E) \propto \left|\sum_{k=1}^{N-1} a_k(E) + Cf(E)\right|^2 = \left|\sum_{k=1}^{N-1} \frac{A_k f_k(E)}{E - M_k} + \frac{A_N f(E)}{E - M_N}\right|^2,$$

where $A_N = -M_N C$, $M_N \to \infty$, and we shall consider it formally as the particular case of interfering resonances.

In the fit of experimental data using the cross section (7), the parameters $A_k$ and $M_k$ are to be determined. The problem of finding equivalent solutions for the parameters is as follows: we must find all such numbers $A_{kx}$ and $M_{kx}$ that at the interval $[E_1, E_2]$ the equality

$$\left|\sum_{k=1}^{N} \frac{A_k f_k(E)}{E - M_k}\right|^2 = \left|\sum_{k=1}^{N} \frac{A_{kx} f_k(E)}{E - M_{kx}}\right|^2 \tag{8}$$

holds.

Let us write this equation in the form

$$\Phi(E)\Phi(E^*)^* = \Phi_x(E)\Phi_x(E^*)^*, \tag{9}$$

where

$$\Phi(E) = \sum_{k=1}^{N} \frac{A_k f_k(E)}{E - M_k}, \qquad \Phi_x(E) = \sum_{k=1}^{N} \frac{A_{kx} f_k(E)}{E - M_{kx}}.$$

The functions in the left- and right-hand sides of (9) are analytic, therefore they coincide not only in the interval $[E_1, E_2]$, but also in the entire region $\Omega$. It follows that, up to numeration, the function $\Phi_x(E)$ has poles $M_{kx} = \{M_k \text{ or } M_k^*\}$. Taking into account that poles $M_{kx}$ should be located in the lower half-plane of a complex variable, we obtain $M_{kx} = M_k$. If there is a background, then its amplitude does not depend on $M_N$ when $M_N \to \infty$, therefore we can also assume $M_{Nx} = M_N$ for it. Thus, the ambiguity in the determination of parameters may be due to parameters $A_k$ only (we assume that the functions $f_k$ are unambiguously related to the poles $M_k$). The problem can therefore be reformulated as follows: for a given set $\mathbf{A} = (A_1, \ldots, A_N)$ (all $A_k \neq 0$) and functions

$$R_k(E) = \frac{f_k(E)}{E - M_k} \quad (k = 1, \ldots, N), \tag{10}$$

we have to find all sets $\mathbf{A}_x = (A_{1x}, \ldots, A_{Nx})$ for which at the interval $E_1 < E < E_2$ the equality

$$\left|\sum_{k=1}^{N} A_k R_k\right| = \left|\sum_{k=1}^{N} A_{kx} R_k\right| \tag{11}$$

holds.

We note first that all solutions, if exist, are defined up to a phase factor $e^{i\varphi}$ ($\varphi$ is a real constant). Therefore, we shall not distinguish the solutions which differ by such a factor. It is obvious that there is always a solution $\mathbf{A}_x = \mathbf{A} e^{i\varphi}$, which we shall call the trivial one. Suppose now that there is a non-trivial solution $\mathbf{A}_x \neq \mathbf{A} e^{i\varphi}$. Then for an arbitrary complex value $E$ we can write

$$\sum_{k=1}^{N} A_k R_k(E) = F(E) \sum_{k=1}^{N} A_{kx} R_k(E), \tag{12}$$

where $|F(E)| = 1$ at $E_1 < E < E_2$. Function $F(E)$ maps the interval $[E_1, E_2]$ of the real axis on an arc of circle $C$ of unit radius centered at point 0. This arc can not be of zero length, because otherwise $F(E) = e^{i\varphi}$, and since the functions $R_k$ are linearly independent we would have $\mathbf{A}_x = \mathbf{A} e^{-i\varphi}$. Let $p(z) = \dfrac{z - \lambda}{z - \lambda^*}$ be a rational function which transforms the real axis into the circle $C$, where $\lambda$ is a complex number, $\lambda \neq \lambda^*$. Since we have identically $F = pp^{-1}F = ph$, it follows that the function $F$ can be written as

$$F = \frac{h - \lambda}{h - \lambda^*}, \tag{13}$$

where $h(E) = p^{-1}F$ is a real function at $E_1 < E < E_2$. Substituting (13) in (12), we get

$$\sum_{k=1}^{N} \alpha_k (h - \beta_k) R_k \equiv 0, \tag{14}$$

where

$$\begin{cases} \alpha_k = A_k - A_{kx}, \\ \beta_k = \dfrac{A_k \lambda^* - A_{kx} \lambda}{A_k - A_{kx}}. \end{cases} \tag{15}$$

Introducing functions $q_k(E) \equiv (h(E) - \beta_k) R_k(E)$, we obtain that $R_k$ can be written as

$$R_k(E) = \frac{q_k(E)}{h(E) - \beta_k}, \tag{16}$$

where functions $q_k(E)$ are linearly dependent: $\sum_{k=1}^{N} \alpha_k q_k \equiv 0$, $\sum_{k=1}^{N} |\alpha_k| > 0$.

Conversely, suppose that functions $R_k$ can be represented in the form (16). From the relations (15) we see that for a given $\lambda$ parameters $A_k$ and $A_{kx}$ are uniquely determined by the quantities $\alpha_k$, $\beta_k$

$$\begin{cases} A_k = \alpha_k \dfrac{\lambda - \beta_k}{\lambda - \lambda^*}, \\ A_{kx} = \alpha_k \dfrac{\lambda^* - \beta_k}{\lambda - \lambda^*}. \end{cases} \tag{17}$$

It follows that for given functions $R_k$ of the form (16), for some values of parameters, namely

$$A_k = \alpha_k \frac{\lambda - \beta_k}{\lambda - \lambda^*}, \tag{18}$$

where $\lambda$ is an arbitrary complex number, $\lambda \neq \lambda^*$, a non-trivial solution

$$A_{kx} = \alpha_k \frac{\lambda^* - \beta_k}{\lambda - \lambda^*} = A_k \frac{\lambda^* - \beta_k}{\lambda - \beta_k} \tag{19}$$

exists. Indeed, for such $\mathbf{A}$ and $\mathbf{A}_x$ the relations (15) holds. Since functions $q_k(E)$ are linearly dependent, the identity (14) is also held. Substituting (15) in (14), we get

$$\sum_{k=1}^{N} A_k R_k(E) = \frac{h - \lambda}{h - \lambda^*} \sum_{k=1}^{N} A_{kx} R_k(E).$$

It follows that for $E_1 < E < E_2$, when $h(E)$ is real, the equation (11) holds.

For given functions $R_k$ of the form (16) the allowed values of $A_k$, for which a non-trivial solution exists, are located on some two-dimensional surface, because they depend on two parameters: $\lambda$ and an arbitrary factor by which $\alpha_k$ may be multiplied.

Thus, we have the following

**Theorem.** *The equation*

$$\left|\sum_{k=1}^{N} A_k R_k(E)\right| = \left|\sum_{k=1}^{N} A_{kx} R_k(E)\right|$$

*in the interval $E_1 < E < E_2$ has a non-trivial solution $\mathbf{A}_x \neq \mathbf{A}e^{i\varphi}$ if and only functions $R_k(E)$ can be represented in the form*

$$R_k(E) = \frac{q_k(E)}{h(E) - \beta_k},$$

*where $\beta_k$ are complex numbers, $h(E)$ is a real function at $E_1 < E < E_2$, and $q_k(E)$ are linearly dependent functions: $\sum_{k=1}^{N} \alpha_k q_k \equiv 0$ for some numbers $\alpha_k$, $\sum_{k=1}^{N} |\alpha_k| > 0$. Herewith $A_k$ and $A_{kx}$ are related to $\alpha_k$ and $\beta_k$ as follows:*

$$\begin{cases} A_k = \alpha_k \dfrac{\lambda - \beta_k}{\lambda - \lambda^*}, \\ A_{kx} = \alpha_k \dfrac{\lambda^* - \beta_k}{\lambda - \lambda^*} = A_k \dfrac{\lambda^* - \beta_k}{\lambda - \beta_k}, \end{cases} \qquad \begin{cases} \alpha_k = A_k - A_{kx}, \\ \beta_k = \dfrac{A_k \lambda^* - A_{kx} \lambda}{A_k - A_{kx}}, \end{cases} \qquad (20)$$

*where $\lambda$ is a complex number, $\lambda \neq \lambda^*$.*

Let us consider consequences of this theorem.

Firstly, it is easy to find the solutions for interfering BW-resonances. The cross section in this case is

$$\sigma(E) \propto \left|\sum_{k=1}^{N} \frac{A_k}{E - M_k}\right|^2,$$

and functions $R_k$ already have the form (16), where $h(E) = E$, functions $q_k(E) \equiv 1$ are linear dependent, and constants $\beta_k$ equal $\beta_k = M_k = E_{Rk} - i\Gamma_k/2$ $(k = 1,\ldots,N)$. The condition of linear dependence for functions $q_k(E)$ looks like $\sum_{k=1}^{N} \alpha_k = 0$. Expressing $\alpha_k$ from (20) through $A_k$, $M_k$ and $\lambda$, we can write this condition in the form

$$\sum_{k=1}^{N} \alpha_k = \sum_{k=1}^{N} A_k \frac{\lambda - \lambda^*}{\lambda - M_k} = 0,$$

or

$$\sum_{k=1}^{N} A_k \frac{1}{\lambda - M_k} = \sum_{k=1}^{N} A_k R_k(\lambda) = \Phi(\lambda) = 0.$$

It follows that if the function $\Phi(E)$ has zeros which do not lie on the real axis, then a non-trivial solution for any set of parameters $A_k$ exists. Indeed, if $\lambda$ ($\lambda \neq \lambda^*$) is the root of the equation $\Phi(E) = 0$, which we shall call the characteristic equation, then any such numbers $A_k$ can be represented in the form (20):

$$A_k = \alpha_k \frac{\lambda - M_k}{\lambda - \lambda^*},$$

where

$$\alpha_k = A_k \frac{\lambda - \lambda^*}{\lambda - M_k}, \quad \sum_{k=1}^{N} \alpha_k = 0.$$

Then the non-trivial solution, according to (20), is equal to

$$A_{kx} = A_k \frac{\lambda^* - M_k}{\lambda - M_k}.$$

In this case the function $\Phi(E)$ is a proper rational function,

$$\Phi(E) = \frac{c_N E^{N-1} + c_{N-1} E^{N-2} + \ldots + c_1}{(E - M_1)(E - M_2)\ldots(E - M_N)} = \frac{c_{P+1}(E - \lambda_1)(E - \lambda_2)\ldots(E - \lambda_P)}{(E - M_1)(E - M_2)\ldots(E - M_N)}, \quad (21)$$

where $\lambda_k$ $(k = 1,\ldots,P)$ are the roots of the characteristic equation, $0 \le P \le N-1$. Herewith the coefficients $c_k$ $(k = 1,\ldots,N)$ are related to the parameters $A_k$ as follows: $c_k = \sum_{n=1}^{N} M_{kn} A_n$, where

$$M_{kn}^{(N)} = (-1)^{N-k} \sum_{\substack{i_1 < i_2 < \ldots < i_{N-k}, \\ i_l \ne n}} M_{i_1} M_{i_2} \ldots M_{i_{N-k}}, \quad M_{Nn}^{(N)} = 1, \quad (k = 1,2,\ldots,N-1;\ n = 1,2,\ldots,N)$$

are the elements of the $N \times N$ matrix (we shall call it the mass matrix) $\mathbf{M}^{(N)}$,

$$\mathbf{M}^{(N)} = \begin{pmatrix} (-1)^{N-1} M_2 M_3 \cdots M_N & (-1)^{N-1} M_1 M_3 \cdots M_N & \cdots & (-1)^{N-1} M_1 M_2 \cdots M_{N-1} \\ \ldots & \ldots & \ldots & \ldots \\ M_2 M_3 + M_2 M_4 + \ldots + M_{N-1} M_N & M_1 M_3 + M_1 M_4 + \ldots + M_{N-1} M_N & \cdots & M_1 M_2 + M_1 M_3 + \ldots + M_{N-2} M_{N-1} \\ -M_2 - M_3 - \ldots - M_N & -M_1 - M_3 - \ldots - M_N & \cdots & -M_1 - M_2 - \ldots - M_{N-1} \\ 1 & 1 & \cdots & 1 \end{pmatrix}.$$

Successively taking all $P$ roots, we obtain the solutions

$$A_{kx}^{(n)} = A_k \frac{\lambda_n^* - M_k}{\lambda_n - M_k} \quad (n = 1,2,\ldots,P). \quad (22)$$

The relation (12) for these solutions has the following form:

$$\Phi(E) = \frac{E - \lambda_n}{E - \lambda_n^*} \Phi_x^{(n)}(E), \quad (23)$$

i.e. $\Phi_x^{(n)}(E)$ differs from $\Phi(E)$ only in that it has a zero $\lambda_n^*$ instead of $\lambda_n$. Similarly, taking another zero $\lambda_s$ $(s \ne n)$ of the function $\Phi_x^{(n)}(E)$, one can obtain from $\mathbf{A}_x^{(n)}$ another solution

$$A_{kx} = A_{kx}^{(n)} \frac{\lambda_s^* - M_k}{\lambda_s - M_k} = A_k \frac{\lambda_n^* - M_k}{\lambda_n - M_k} \frac{\lambda_s^* - M_k}{\lambda_s - M_k}.$$

Continuing in this way we obtain $2^P - 1$ solutions

$$A_{kx}^{(n)} = A_k \prod_{j=1}^{r} \frac{\lambda_{n_j}^* - M_k}{\lambda_{n_j} - M_k} \quad (n = 1,2,\ldots,2^P - 1,\ n_1 < n_2 < \ldots < n_r,\ 1 \le r \le P). \quad (24)$$

Taking into account the form (21) of the function $\Phi(E)$, it is obvious that there are no other solutions in this case. Each solution is uniquely determined by a corresponding set of roots of the characteristic equation. Hence the solutions (24) can be written together with the trivial solution in a uniform manner, numbering each of them by its own set $\mathbf{\Lambda} = (\lambda_1', \lambda_2', \ldots, \lambda_P')$, where $\lambda_n' = \{\lambda_n \text{ or } \lambda_n^*\}$ $(n = 1,\ldots,P)$:

$$A_{kx}^{(\mathbf{\Lambda})} = A_k \prod_{n=1}^{P} \frac{\lambda_n' - M_k}{\lambda_n - M_k}. \quad (25)$$

The trivial solution corresponds to the set $\mathbf{\Lambda} = \{\lambda_1, \lambda_2, \ldots, \lambda_P\}$. It is also obvious that there is symmetry between the solutions: each solution can be obtained from any other by suitable complex conjugation of some roots.

If some root of the characteristic equation is real, or it has multiplicity greater than one, or it is complex conjugate to another root, then some of the solutions (25) coincide. Thus, the total number of solutions for interfering Breit-Wigner resonances, including the trivial one, does not exceed $2^P$.

The interference of BW-resonances with a complex constant $C$,

$$\sigma(E) \propto \left| \sum_{k=1}^{N-1} \frac{A_k}{E - M_k} + C \right|^2,$$

can be formally treated, as noted above, as a particular case of interference of BW-resonances, one of which has an infinitely large mass or width. For that we make a substitution $A_N = -CM_N$ and tend $M_N \to \infty$. As a result, we have the same formulae (25) for the solutions, where $\lambda_n$ ($n = 1, \ldots, P$) are the roots of the characteristic equation (21) with coefficients $c_k$ ($k = 1, \ldots, N$), which in this case are

$$\begin{pmatrix} c_1 \\ c_2 \\ \cdots \\ c_{N-1} \\ c_N \end{pmatrix} = \begin{pmatrix} & & & (-1)^{N-1} M_1 M_2 \cdots M_{N-1} \\ & \mathbf{M}^{(N-1)} & & \cdots \\ & & & M_1 M_2 + M_1 M_3 + \ldots + M_{N-2} M_{N-1} \\ & & & -M_1 - M_2 - \ldots - M_{N-1} \\ 0 & \cdots & 0 & 1 \end{pmatrix} \begin{pmatrix} A_1 \\ A_2 \\ \cdots \\ A_{N-1} \\ C \end{pmatrix}.$$

From formulae (25) we see that $C_x^{(\Lambda)} = C$ for all solutions.

Let us consider in more detail the simplest cases of two and three interfering BW-resonances.

In case $N = 2$ the mass matrix is

$$\mathbf{M}^{(2)} = \begin{pmatrix} -M_2 & -M_1 \\ 1 & 1 \end{pmatrix},$$

the characteristic equation (given that $c_2 \ne 0$) has the form

$$\Phi(E) = \frac{c_2 E + c_1}{(E - M_1)(E - M_2)} = \frac{c_2 (E - \lambda_1)}{(E - M_1)(E - M_2)} = 0,$$

and its root equals

$$\lambda_1 = -\frac{c_1}{c_2} = \frac{A_1 M_2 + A_2 M_1}{A_1 + A_2}. \tag{26}$$

If it has a non-zero imaginary part, we have two solutions,

$$\begin{cases} A_{kx}^{(\lambda_1)} = A_k & - \text{trivial solution,} \\ A_{kx}^{(\lambda_1^*)} = A_k \dfrac{\lambda_1^* - M_k}{\lambda_1 - M_k} & (k = 1, 2). \end{cases} \tag{27}$$

Substituting here the expression (26) for the root $\lambda_1$, we obtain expressions for a non-trivial solution through the initial parameters $A_k$, $M_k$:

$$\begin{cases} A_{1x}^{(\lambda_1^*)} = \dfrac{A_1^* (M_1 - M_2^*) + A_2^* (M_1 - M_1^*)}{M_1 - M_2}, \\ A_{2x}^{(\lambda_1^*)} = \dfrac{A_1^* (M_2^* - M_2) + A_2^* (M_1^* - M_2)}{M_1 - M_2}. \end{cases} \tag{28}$$

If we rewrite these formulae using appropriate notations, they coincide with the solution obtained in [5] (formulae (17)–(19)).

If $c_2 = 0$, then there are no non-trivial solutions. We can say that in this case the root $\lambda_1 = -\dfrac{c_1}{c_2}$ goes to infinity, and the solution $\mathbf{A}_x^{(\lambda_1^*)}$ coincides with the trivial one.

When a BW-resonance interferes with a complex constant $C$, the cross section has the form
$$\sigma(E) \propto \left| \frac{A_1}{E - M_1} + C \right|^2,$$
and a non-trivial solution can be obtained from (28) by substitution $A_2 = -M_2 C$, $M_2 \to \infty$:
$$\begin{cases} A_{1x}^{(\lambda_1^*)} = A_1^* + C^*(M_1 - M_1^*), \\ C_x^{(\lambda_1^*)} = C^*. \end{cases}$$
It coincides (again, up to notations) with the solution of Ref. [5] (formulae (23)). In this case the characteristic equation is of the form
$$\frac{C(E - \lambda_1)}{E - M_1} = 0,$$
therefore there is a nontrivial solution only if the root $\lambda_1 = M_1 - A_1/C$ has a non-zero imaginary part.

When $N = 3$ the mass matrix is
$$\mathbf{M}^{(3)} = \begin{pmatrix} M_2 M_3 & M_1 M_3 & M_1 M_2 \\ -M_2 - M_3 & -M_1 - M_3 & -M_1 - M_2 \\ 1 & 1 & 1 \end{pmatrix},$$
and the characteristic equation has the form
$$\Phi(E) = \frac{c_3 E^2 + c_2 E + c_1}{(E - M_1)(E - M_2)(E - M_3)} = \frac{c_3 (E - \lambda_1)(E - \lambda_2)}{(E - M_1)(E - M_2)(E - M_3)} = 0. \tag{29}$$

Given that $c_3 \neq 0$ its roots are equal to $\lambda_{1,2} = \dfrac{-c_2 \pm \sqrt{c_2^2 - 4c_3 c_1}}{2c_3}$. In general case there are four solutions:
$$\begin{cases} 1.\ A_{kx}^{(\lambda_1, \lambda_2)} = A_k - \text{trivial solution}, \\ 2.\ A_{kx}^{(\lambda_1^*, \lambda_2)} = A_k \dfrac{\lambda_1^* - M_k}{\lambda_1 - M_k}, \\ 3.\ A_{kx}^{(\lambda_1, \lambda_2^*)} = A_k \dfrac{\lambda_2^* - M_k}{\lambda_2 - M_k}, \\ 4.\ A_{kx}^{(\lambda_1^*, \lambda_2^*)} = A_k \dfrac{\lambda_1^* - M_k}{\lambda_1 - M_k} \dfrac{\lambda_2^* - M_k}{\lambda_2 - M_k} \quad (k = 1,2,3). \end{cases} \tag{30}$$

If some of the roots are real, the number of solutions is reduced. For example, if $\lambda_1 = \lambda_1^*$, then there is only one non-trivial solution, $\mathbf{A}_x^{(\lambda_1, \lambda_2^*)} = \mathbf{A}_x^{(\lambda_1^*, \lambda_2^*)}$. If $\lambda_2 = \lambda_2^*$, then there is also one non-trivial solution $\mathbf{A}_x^{(\lambda_1^*, \lambda_2)} = \mathbf{A}_x^{(\lambda_1^*, \lambda_2^*)}$ only. If $\lambda_1 = \lambda_1^*$ and $\lambda_2 = \lambda_2^*$, then there are no non-trivial solutions. Also, the number of solutions is reduced when the root is multiple, $\lambda_1 = \lambda_2$, or when $\lambda_1 = \lambda_2^*$. In the former case there are two non-trivial solutions, $\mathbf{A}_x^{(\lambda_1^*, \lambda_2)} = \mathbf{A}_x^{(\lambda_1, \lambda_2^*)}$ and $\mathbf{A}_x^{(\lambda_1^*, \lambda_2^*)}$, while in the latter case there are two: $\mathbf{A}_x^{(\lambda_1^*, \lambda_2)}$ and $\mathbf{A}_x^{(\lambda_1, \lambda_2^*)}$.

If $c_3 = 0$ but $c_2 \neq 0$ (i.e. one of roots is located at infinity), then instead of (29) we get equation

$$\Phi(E) = \frac{c_2 E + c_1}{(E-M_1)(E-M_2)(E-M_3)} = \frac{c_2(E-\lambda_1)}{(E-M_1)(E-M_2)(E-M_3)} = 0,$$

where root $\lambda_1$ equals $\lambda_1 = -\frac{c_1}{c_2}$. If it has a nonzero imaginary part, then there is an unique non-trivial solution

$$A_{kx}^{(\lambda_1^*)} = A_k \frac{\lambda_1^* - M_k}{\lambda_1 - M_k} \quad (k=1,2,3).$$

If $c_3 = c_2 = 0$ (both roots are at infinity), then there are no non-trivial solutions.

Making in the cross section the substitution $A_3 = -CM_3$, $M_3 \to \infty$, we obtain as a particular case the interference of two BW-resonances with a background $C$:

$$\sigma(E) \propto \left| \frac{A_1}{E-M_1} + \frac{A_2}{E-M_2} + C \right|^2,$$

Solutions are given by formulae (30), where the roots $\lambda_{1,2} = \frac{-c_2 \pm \sqrt{c_2^2 - 4c_3 c_1}}{2c_3}$ are calculated with the coefficients $c_k$ $(k=1,2,3)$ which are equal to

$$\begin{pmatrix} c_1 \\ c_2 \\ c_3 \end{pmatrix} = \begin{pmatrix} -M_2 & -M_1 & M_1 M_2 \\ 1 & 1 & -M_1 - M_2 \\ 0 & 0 & 1 \end{pmatrix} \begin{pmatrix} A_1 \\ A_2 \\ C \end{pmatrix}.$$

Special cases, when the number of solutions decreases, are considered in the same way as for three resonances.

Note also that although we have considered above the Breit-Wigner resonances with amplitudes (1), it is easy to generalize that for the relativistic form (3) of the amplitude – it is sufficient to use the variable $s$ instead of $E$ and the elements $M_k = m_k^2 - im_k\Gamma_k$ instead of $M_k = E_{Rk} - i\Gamma_k/2$.

Now we turn to a more general form of the resonances (10). Consider first the case of two interfering resonances, $N = 2$. If for a given set of parameters $\mathbf{A}$ a non-trivial solution $\mathbf{A}_x$ exists, then, according to the theorem, the functions $q_k$ in (16) are proportional to each other: $q_2 = \alpha q_1$. Therefore

$$\Phi(E) = \sum_{k=1}^{2} A_k R_k = q_1 \left( A_1 \frac{1}{h-\beta_1} + A_2 \frac{\alpha}{h-\beta_2} \right), \qquad (31)$$

and introducing a new variable $z = h(E)$, we see that resonances have the Breit-Wigner form (up to the function $q_1$ which is inessential for interference). Thus, if a non-trivial solution $\mathbf{A}_x$ exists, then it is unique.

A necessary and sufficient condition for the existence of a non-trivial solution in this case can also be formulated in a different way. Indeed, the theorem implies that for the case $N = 2$ the non-trivial solution exists if and only if the ratio of the functions $R_1$ and $R_2$ is equal to

$$R \equiv \frac{R_2}{R_1} = \alpha \frac{h-\beta_1}{h-\beta_2}, \qquad (32)$$

i.e. the function $R$ maps the interval $[z_1 = h(E_1), z_2 = h(E_2)]$ of the real axis on an arc of the circle in a complex plane – the result obtained also in Ref. [4]. Thus, in the case of two interfer-ing resonances it's sufficient to check whether the mapping $R$ is an arc of a circle. If so, the non-trivial solution exists and the representation (16) can be found writing $R$ in the form $R = pp^{-1}R = ph$, where $p(z) = \alpha \dfrac{z - \beta_1}{z - \beta_2}$ is a rational function which transforms the real axis into this circle. Further, this solution can be found by using the formulae (28) for the BW-resonances:

$$\begin{cases} A_{1x} = \dfrac{A_1^*(\beta_1 - \beta_2^*) + \alpha^* A_2^*(\beta_1 - \beta_1^*)}{\beta_1 - \beta_2}, \\ \alpha A_{2x} = \dfrac{A_1^*(\beta_2^* - \beta_2) + \alpha^* A_2^*(\beta_1^* - \beta_2)}{\beta_1 - \beta_2}. \end{cases} \qquad (33)$$

Let us consider another limiting case of these formulae. We substitute in (31) $A_2 = -C\beta_2$, where $C$ is a complex constant, and tend $\beta_2$ to infinity. In this case

$$\Phi(E) \to A_1 \dfrac{q_1}{h - \beta_1} + C\alpha q_1 = A_1 R_1 + Cf,$$

i.e. we obtain the interference of a resonance with the background, which amplitude $Cf = C\alpha q_1$ does not have poles near the interval $[E_1, E_2]$. A circle (32) wherein degenerates into a straight line:

$$\widetilde{R} \equiv -\beta_2 R = -\beta_2 \alpha \dfrac{h - \beta_1}{h - \beta_2} \to \alpha(h - \beta_1).$$

Thus, we find that when the resonance interferes with the background,

$$\sigma(E) \propto |A_1 R_1(E) + Cf(E)|^2,$$

the non-trivial solution exists if and only if the function $\widetilde{R} = \dfrac{f}{R_1}$ maps the interval $[E_1, E_2]$ of real axis on an interval of the straight line $p(E) = \alpha(E - \beta_1)$ in a complex plane. This solution can be found by taking the limit $\beta_2 \to \infty$ in formulae (33):

$$\begin{cases} A_{1x} = A_1^* + \alpha^* C^*(\beta_1 - \beta_1^*), \\ C_x = C^* \dfrac{\alpha^*}{\alpha}. \end{cases}$$

For interfering resonances of the general form (10) at $N \geq 3$ such arguments are not applicable, so we restrict ourselves to the case of weak dependence of functions $f_k$ on $E$, i.e. we shall assume that the resonances are of "almost Breit-Wigner" form. From the above results it follows that if Breit-Wigner resonances with parameters $\mathbf{A}_0$ interfere, then fitting the experimental data we obtain a local maximum $L$ of a likelihood function with parameters $\mathbf{A}$ close to $\mathbf{A}_0$. Moreover, there will be also local maxima $L_x^{(\mathbf{A})}$ at the parameter values (25) obtained from $\mathbf{A}$, and these maxima have the same height. Other local maxima, if they exist, lie below. Now let us change the cross section (data also are changed) in a continuous manner so that the BW-amplitudes are replaced by amplitudes (6):

$$\dfrac{A_k}{E - M_k} \to \dfrac{A_k f_k(E)}{E - M_k} \quad (k = 1, \ldots, N). \qquad (34)$$

Maxima $L_x^{(\mathbf{A})}$ wherein will be displaced, descending below the maximum $L$, provided that non-trivial solutions of the equation (8) do not exist. If functions $f_k$ depend on $E$ rather weakly, then the parameters of the maxima $L_x^{(\mathbf{A})}$ will also give a good approximation of the original cross section. In principle, during the transition (34) new maxima can also arise, however, if they are lo-cated far away from $L_x^{(\Lambda)}$, they give worst approximation of the cross section. Therefore it is enough to search for solutions for the resonance parameters close to values of the parameters which are related to each other as solutions for BW-resonances. If in the fit of data a solution (i.e. a local maximum of the likelihood function) for parameters is obtained, then approximate values of other possible solutions can be found replacing functions $f_k$ by constants and using the formulae (25). As such constants it's natural to take $f_k(M_k)$, that corresponds to the replacement of functions $R_k$ in the amplitudes of the resonances by the principal parts of their Laurent expansions near the poles $M_k$. If resonances are given in the form of BW-resonances with widths depending on energy, i.e.

$$A_k R_k(E) = \frac{A_k}{E - m_k + i\Gamma_k(E)/2} = \frac{A_k}{E - \widetilde{M}_k(E)} \quad (k = 1, \ldots, N), \qquad (35)$$

and rather narrow (as well as widths weakly depend on E, which is assumed), then for such replacement it is enough to substitute in (35) the constants $M_k = m_k - i\Gamma_k(m_k)/2$ instead of functions $\widetilde{M}_k(E)$. If close to the calculated values of parameters $A_{kx}^{(\Lambda)}$ the solutions are found, then using a criterion $\chi^2$ we can choose among the entire solutions the true one. When the criterion $\chi^2$ does not allow a clear choice in favor of a particular parameter values, it is necessary to consider additional (e.g., theoretical) arguments.

The locations of maxima $L_x^{(\Lambda)}$ relative to the main maximum $L$ can be found more accurately in a different way. For this we note that relations (23), from which the solutions (25) for BW-resonances are obtained, are fulfilled identically on $A_k$ provided that $A_{kx}^{(n)}$ are expressed in terms of $A_k$ according to (22). This means that if the constants $A_k$ in (25) are replaced by arbitrary functions $\widetilde{A}_k(E)$, then for functions

$$\widetilde{A}_{kx}^{(\Lambda)}(E) = \widetilde{A}_k(E) \prod_{n=1}^{N-1} \frac{\lambda_n'(E) - M_k}{\lambda_n(E) - M_k} \qquad (36)$$

in the interval $[E_1, E_2]$ the equality

$$\left| \sum_{k=1}^{N} \frac{\widetilde{A}_k(E)}{E - M_k} \right| = \left| \sum_{k=1}^{N} \frac{\widetilde{A}_{kx}^{(\Lambda)}(E)}{E - M_k} \right|. \qquad (37)$$

holds. In (36) functions $\lambda_n(E)$ are expressed through $c_k(E) = \sum_{n=1}^{N} M_{kn} \widetilde{A}_n(E)$ in the same way as for BW-resonances roots $\lambda_n$ are expressed through $c_k = \sum_{n=1}^{N} M_{kn} A_n$ (for example, for two resonances $\lambda_1(E) = -c_1(E)/c_2(E)$), and $\lambda_n'(E) = \{\lambda_n(E) \text{ or } \lambda_n(E^*)^*\}$ (here for $\lambda_n'(E)$ we take $\lambda_n(E^*)^*$ instead of $\lambda_n(E)^*$, because in the interval $[E_1, E_2]$ the equation (37) is satisfied for any choice, but functions $\lambda_n(E^*)^*$ are more convenient since for them $\widetilde{A}_{kx}^{(\Lambda)}(E)$ are analytic functions). The functions $\widetilde{A}_{kx}^{(\Lambda)}(E)$ in general depend on $E$ differently than $\widetilde{A}_k(E)$. Therefore, if we write the resonance amplitudes in the form

$$\frac{A_k f_k(E)}{E - M_k} = \frac{\widetilde{A}_k(E)}{E - M_k},$$

then the functions (36) do not give the solutions, since they can not be represented as $\widetilde{A}_{kx}^{(\Lambda)}(E) = A_{kx}^{(\Lambda)} f_k(E)$, where $A_{kx}^{(\Lambda)}$ are some constants. However, if we replace the functions $\widetilde{A}_{kx}^{(\Lambda)}(E)$ in the right hand side of (37) by the functions $\dfrac{\widetilde{A}_{kx}^{(\Lambda)}(M_k)}{f_k(M_k)} f_k(E)$ then we obtain

$$\sum_{k=1}^{N} \frac{\widetilde{A}_{kx}^{(\Lambda)}(E)}{E - M_k} \approx \sum_{k=1}^{N} \frac{\widetilde{A}_{kx}^{(\Lambda)}(M_k)}{f_k(M_k)} \frac{f_k(E)}{E - M_k}, \qquad (38)$$

because the principal parts (which mainly determine these functions) of Laurent expansions of functions about the poles $M_k$ in left- and right-hand sides of (38) coincide. Therefore the values of parameters

$$A_{kx}^{(\Lambda)} = \frac{\widetilde{A}_{kx}^{(\Lambda)}(M_k)}{f_k(M_k)} \qquad (39)$$

will give approximate positions of maxima $L_x^{(\Lambda)}$.

Similar arguments are also applicable when resonances are given in the form (35). The values of parameters, where maxima $L_x^{(\Lambda)}$ are located, approximately equal

$$A_{kx}^{(\Lambda)} = \widetilde{A}_{kx}^{(\Lambda)}(M_k), \qquad (40)$$

where $M_k$ are the corresponding resonant poles, and functions

$$\widetilde{A}_{kx}^{(\Lambda)}(E) = A_k \prod_{n=1}^{N-1} \frac{\lambda_n'(E) - \widetilde{M}_k(E)}{\lambda_n(E) - \widetilde{M}_k(E)} \qquad (41)$$

are evaluated analogously to the case of BW-resonances, using the functions $\widetilde{M}_k(E)$ instead of parameters $M_k$.

## 3. Numerical experiments

Let us illustrate the obtained results by numerical simulation. We consider several examples of fitting data for the reaction when three resonances with energy-dependent widths interfere. We take the cross section, which was considered in [5]:

$$\sigma(s) = \frac{m_1^4}{s^2} \left(\frac{s^2 - 4\mu^2}{m_1^2 - 4\mu^2}\right)^{\frac{3}{2}} \left|\sum_{k=1}^{3} \frac{2m_k \alpha_k e^{i\psi_k}}{s - m_k^2 + i\Gamma_k m_k g_k(s)}\right|^2, \quad g_k(s) = \left(\frac{s - 4\mu^2}{m_k^2 - 4\mu^2}\right)^{\frac{3}{4}}, \qquad (42)$$

with the parameters:

$$m_1 = 782.6, \ m_2 = 1019.4, \ m_3 = 1200, \ \Gamma_1 = \Gamma_2 = \Gamma_3 = 100, \ \mu = 350,$$
$$\alpha_1 = \alpha_2 = 10, \ \alpha_3 = 3, \ \psi_3 = 0°, \ \psi_2 = 155°, \ \psi_3 = 30°. \qquad (43)$$

In Fig. 1 on the left this cross section is shown, as well as values of "measurements" of cross section, obtained in a numerical experiment. 100 such experiments with the statistics of $10^6$ events were done. The resulting "experimental" points were fitted by the cross section (42) with all parameters except for $\mu$ varying freely. The fit was performed by the maximum likelihood method using the package MINUIT [6]. As a result, in each experiment usually four local maxima were found. The average values of parameters for each maximum are shown in Table 1. One can see that the first found maximum $L_1$ corresponds to the initial set of parameters (43). On the right in Fig. 1 the two-dimensional plot for values $\chi_n^2$ ($n = 2,3,4$) of function $\chi^2$ in the other three local maxima $L_2$, $L_3$, $L_4$, versus $\chi_1^2$, is shown. The third maximum in most cases differs by $\chi^2$ from the first one by less than unity, i.e. the first and third solutions for the resonance parameters are experimentally indistinguishable, or "degenerate", to use the terminology of Ref. [5]. Thus, in this case the statistics of the experiment are not sufficient to uniquely identify the true solution.

Table 1. The average values of the parameters of resonances, obtained in 100 numerical experiments, in the local maxima of the likelihood function for the cross section (42) with parameters (43).

|       | $m_{1x}$ | $m_{2x}$ | $m_{3x}$ | $\Gamma_{1x}$ | $\Gamma_{2x}$ | $\Gamma_{3x}$ | $\alpha_{1x}$ | $\alpha_{2x}$ | $\alpha_{3x}$ | $\psi_{2x}-\psi_{1x}$ | $\psi_{3x}-\psi_{1x}$ |
|---|---|---|---|---|---|---|---|---|---|---|---|
| $L_1$ | 782.6 | 1019.4 | 1200.0 | 100.0 | 100.0 | 99.9  | 10.0 | 10.0 | 3.0  | 155.1° | 30.1°  |
| $L_2$ | 783.2 | 1019.6 | 1202.8 | 99.5  | 100.8 | 107.2 | 11.8 | 11.8 | 3.6  | 206.2° | 107.8° |
| $L_3$ | 782.6 | 1019.4 | 1200.0 | 100.1 | 100.0 | 99.7  | 9.8  | 8.9  | 0.63 | 141.7° | 75.2°  |
| $L_4$ | 783.2 | 1019.6 | 1202.7 | 99.5  | 100.7 | 106.9 | 11.6 | 10.5 | 0.89 | 193.5° | 158.3° |

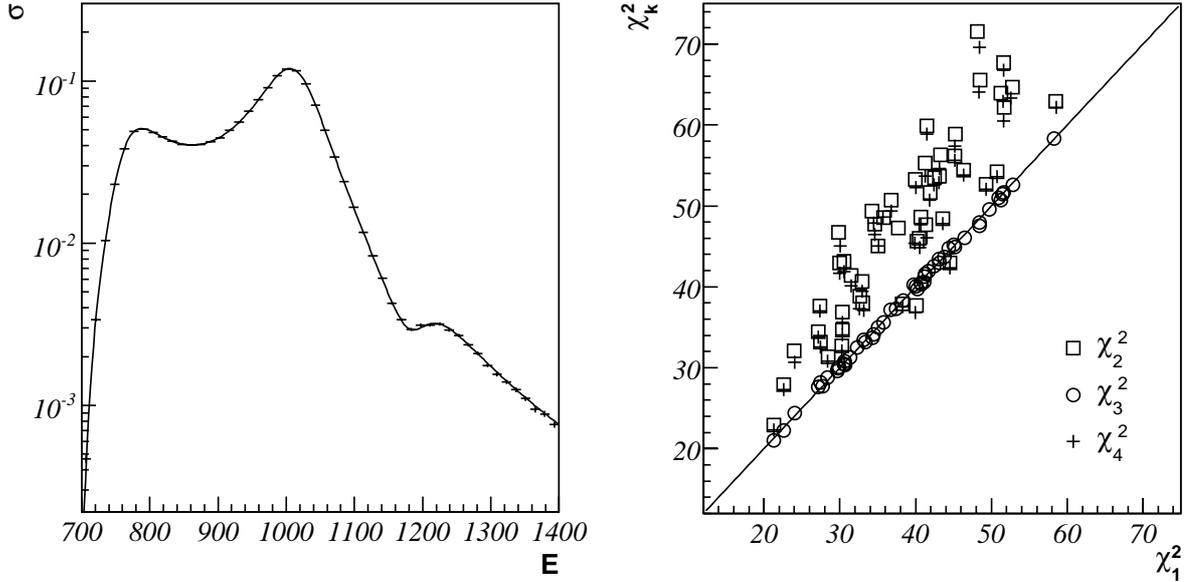

Figure 1. Left: the fit of "experimental" data for cross section (42) with parameters (43). Right: the values $\chi_n^2$ of function $\chi^2$ in local maxima $L_n$ ($n=2,3,4$) versus the value $\chi_1^2$ in local maximum $L_1$, for 50 numerical experiments with statistics of $10^6$ events.

One can understand why local maxima $L_2 - L_4$ have such values of parameters considering the functions $\widetilde{A}_{kx}^{(\Lambda)}$, which, according to (41), are

$$\begin{cases} 1.\ \widetilde{A}_{kx}^{(\lambda_1,\lambda_2)}(s) = A_k, \\ 2.\ \widetilde{A}_{kx}^{(\lambda_1^*,\lambda_2)}(s) = A_k \dfrac{\lambda_1(s)^* - \widetilde{M}_k(s)}{\lambda_1(s) - \widetilde{M}_k(s)}, \\ 3.\ \widetilde{A}_{kx}^{(\lambda_1,\lambda_2^*)}(s) = A_k \dfrac{\lambda_2(s)^* - \widetilde{M}_k(s)}{\lambda_2(s) - \widetilde{M}_k(s)}, \\ 4.\ \widetilde{A}_{kx}^{(\lambda_1^*,\lambda_2^*)}(s) = A_k \dfrac{\lambda_1(s)^* - \widetilde{M}_k(s)}{\lambda_1(s) - \widetilde{M}_k(s)} \dfrac{\lambda_2(s)^* - \widetilde{M}_k(s)}{\lambda_2(s) - \widetilde{M}_k(s)} \quad (k=1,2,3), \end{cases} \qquad (44)$$

where

$$A_k = 2m_k \alpha_k e^{i\psi_k}, \quad \widetilde{M}_k(s) = m_k^2 - i\Gamma_k m_k g_k(s).$$

Turning from the variable s to $E = \sqrt{s}$ we write (keeping the previous notations for functions) functions $\widetilde{A}_{kx}^{(\Lambda)}$ of (44) in the form $\widetilde{A}_{kx}^{(\Lambda)}(E) = 2m_k \alpha_k^{(\Lambda)}(E) \exp(i\psi_k^{(\Lambda)}(E))$, ($k=1,2,3$). Figure 2 shows the real and imaginary parts of "roots" $\lambda_1(E)$, $\lambda_2(E)$, functions $\alpha_k^{(\Lambda)}(E)$, $\alpha_k^{(\Lambda)}(E)$, $\psi_2^{(\Lambda)}(E) - \psi_1^{(\Lambda)}(m_1)$, $\psi_3^{(\Lambda)}(E) - \psi_1^{(\Lambda)}(m_1)$, as well as the average values of parameters in local maxima obtained in fits. Since the resonances are quite narrow, then, in accordance with the formulae (40), the values of the parameters in maxima $L_n$ ($n = 2,3,4$) approximately equal

$$A_{kx}^{(\Lambda)} \approx \widetilde{A}_{kx}^{(\Lambda)}(m_k),$$

where $\Lambda = (\lambda_1^*, \lambda_2), (\lambda_1, \lambda_2^*), (\lambda_1^*, \lambda_2^*)$ for $n = 2, 3, 4$, respectively.

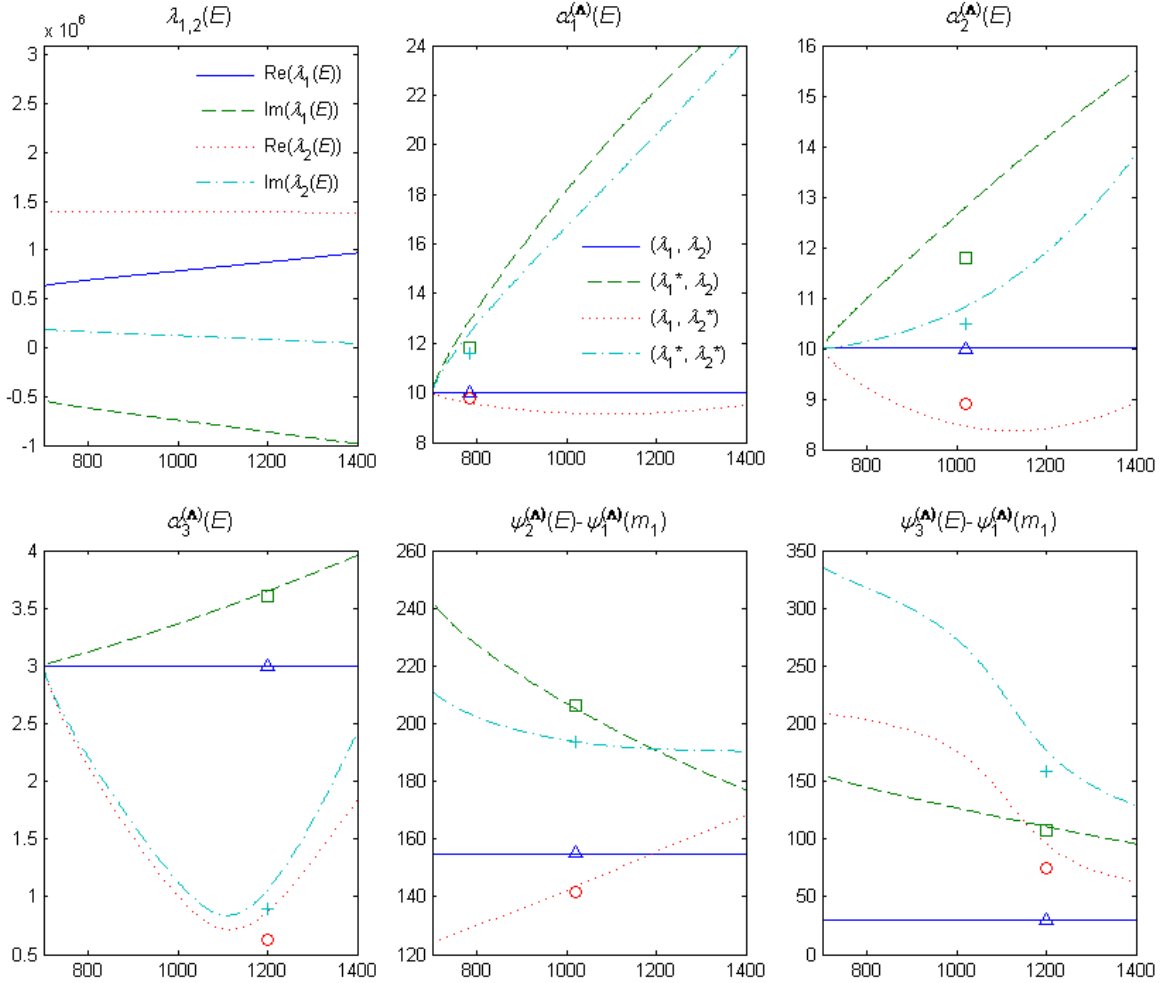

Figure 2. "Roots" $\lambda_1(E)$, $\lambda_2(E)$ and functions $\widetilde{A}_{kx}^{(\Lambda)}(E)$ for the cross section (42) with parameters (43). The markers show average values of parameters in local maxima of the likelihood function: $\triangle - L_1$, $\circ - L_2$, $\square - L_3$, $+ - L_4$.

Functions $\alpha_k^{(\Lambda)}(E)$ ($k = 1,2,3$) merge at the threshold value $E = 2\mu = 700$, since at such $E$ widths of resonances equal zero and functions $\widetilde{M}_k(E)$ are real. Therefore, the more to the right is the resonance, the more additional solutions may differ modulo compared to the initial one. In addition, since the "root" $\lambda_2(E)$ is close to the real axis, we can expect that the third local maximum, corresponding to the solution $\widetilde{\mathbf{A}}_x^{(\lambda_1, \lambda_2^*)}(E)$ will be located closer to the main maximum $L_1$ than the other two. However, one can see from the figure that the parameter $\alpha_{3x}$ for this maximum, on the contrary, differs stronger than others. This is due to the fact that for these values of the initial parameters accidentally $\lambda_2(E) \approx M_3^* \approx m_3^2 + i\Gamma_3 m_3$, therefore, the function $\alpha_3^{(\lambda_1, \lambda_2^*)}(E)$, according to (44), is small for $E \approx m_3$.

Let us change the initial parameters so that the "roots" $\lambda_1(E)$ and $\lambda_2(E)$ are closer to each other. For this we take in the cross section (42) the parameters

$$m_1 = 782.6, \quad m_2 = 1019.4, \quad m_3 = 1200, \quad \Gamma_1 = \Gamma_2 = \Gamma_3 = 100, \quad \mu = 350,$$
$$\alpha_1 = 8.54, \quad \alpha_2 = 4.25, \quad \alpha_3 = 2.85, \quad \psi_3 = 0°, \quad \psi_2 = 92.2°, \quad \psi_3 = 149°. \quad (45)$$

For these values of the parameters the functions $\lambda_1(E)$ and $\lambda_2(E)$ are almost equal at energy $E \approx m_2$. Again in fits the four solutions for the parameters are found with the average values shown in Table 2. The results of numerical simulation are shown in Figures 3, 4. It can be seen that in this case the second and third solutions have values close to the parameter $\alpha_{2x}$, which obviously corresponds to the equality of functions $\widetilde{\mathbf{A}}_x^{(\lambda_1^*, \lambda_2)}$ and $\widetilde{\mathbf{A}}_x^{(\lambda_1, \lambda_2^*)}$ in (44) at equality of "roots".

Table 2. The average values of the parameters of resonances, obtained in 100 numerical experiments, in the local maxima of the likelihood function for the cross section (42) with parameters (45).

|  | $m_{1x}$ | $m_{2x}$ | $m_{3x}$ | $\Gamma_{1x}$ | $\Gamma_{2x}$ | $\Gamma_{3x}$ | $\alpha_{1x}$ | $\alpha_{2x}$ | $\alpha_{3x}$ | $\psi_{2x} - \psi_{1x}$ | $\psi_{3x} - \psi_{1x}$ |
|---|---|---|---|---|---|---|---|---|---|---|---|
| $L_1$ | 782.6 | 1019.4 | 1200.0 | 100.0 | 100.0 | 100.0 | 8.54 | 4.25 | 2.85 | 92.2° | 148.9° |
| $L_2$ | 783.3 | 1019.8 | 1199.5 | 100.2 | 100.8 | 100.7 | 10.1 | 5.98 | 3.69 | 170.7° | 287.1° |
| $L_3$ | 782.9 | 1019.6 | 1199.8 | 100.1 | 100.3 | 100.3 | 9.39 | 7.09 | 4.59 | 161.8° | 352.0° |
| $L_4$ | 783.7 | 1020.1 | 1199.2 | 100.1 | 101.3 | 101.2 | 11.1 | 9.96 | 5.97 | 236.9° | 130.6° |

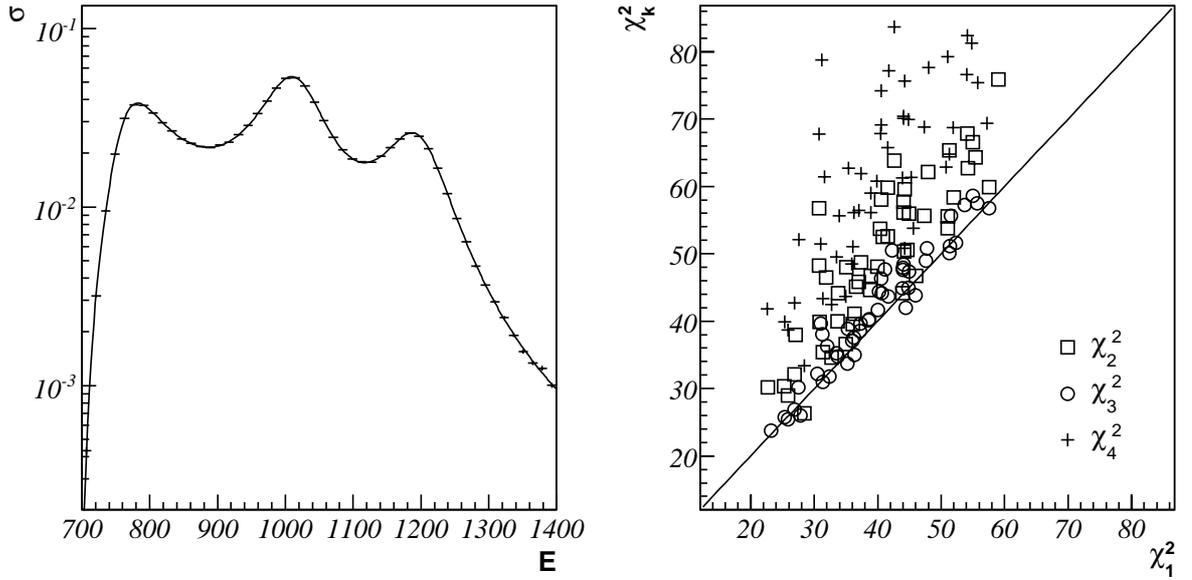

Figure 3. Left: the fit of "experimental" data for cross section (42) with parameters (45). Right: the values $\chi_n^2$ of function $\chi^2$ in local maxima $L_n$ ($n = 2,3,4$) versus the value $\chi_1^2$ in local maximum $L_1$, for 50 numerical experiments with statistics of $10^6$ events.

Let us also consider the case when one of the "roots" $\lambda_k(E)$ "goes to infinity". We take in the cross section the following values of parameters:
$$m_1 = 782.6, \quad m_2 = 1019.4, \quad m_3 = 1200, \quad \Gamma_1 = \Gamma_2 = \Gamma_3 = 100, \quad \mu = 350,$$
$$\alpha_1 = 10, \quad \alpha_2 = 10\frac{m_1}{m_2}, \quad \alpha_3 = 10\frac{m_1}{m_3}, \quad \psi_3 = 0°, \quad \psi_2 = 120°, \quad \psi_3 = 250°. \quad (46)$$

In this case one of "roots" becomes much larger than the characteristic mass scale, $|\lambda_1(E)| >> |\widetilde{M}_k(E)| \approx m_k^2$ ($k = 1,2,3$). Again we have four solutions; the average values of parameters for them are given in Table 3. The results of numerical simulations are shown in Figures 5 and 6. It can be seen that in this case there are two pairs of solutions with close parame-

ters, similarly to the case of interfering BW-resonances only two solutions (one of which is trivial) remain when one of roots goes to infinity.

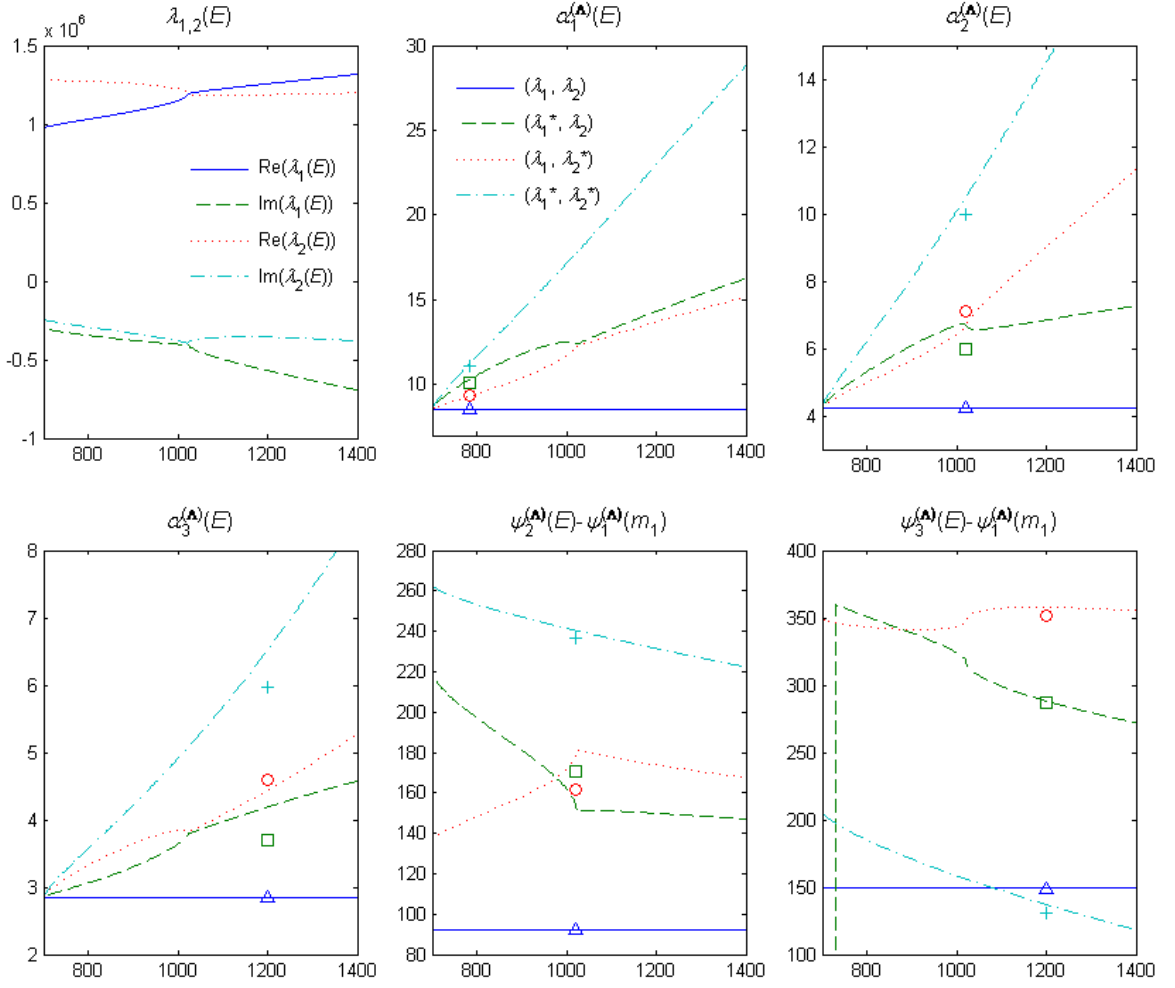

Figure 4. "Roots" $\lambda_1(E)$, $\lambda_2(E)$ and functions $\widetilde{A}_{kx}^{(\Lambda)}(E)$ for the cross section (42) with parameters (45). The markers show average values of parameters in local maxima of the likelihood function: $\triangle - L_1$, $\circ - L_2$, $\square - L_3$, $+ - L_4$.

Table 3. The average values of the parameters of resonances, obtained in 100 numerical experiments, in the local maxima of the likelihood function for the cross section (42) with parameters (46).

|       | $m_{1x}$ | $m_{2x}$ | $m_{3x}$ | $\Gamma_{1x}$ | $\Gamma_{2x}$ | $\Gamma_{3x}$ | $\alpha_{1x}$ | $\alpha_{2x}$ | $\alpha_{3x}$ | $\psi_{2x} - \psi_{1x}$ | $\psi_{3x} - \psi_{1x}$ |
|-------|----------|----------|----------|---------------|---------------|---------------|---------------|---------------|---------------|--------------------------|--------------------------|
| $L_1$ | 782.6    | 1019.4   | 1200.0   | 100.0         | 100.0         | 100.0         | 10.0          | 7.7           | 6.5           | 119.9°                   | 249.7°                   |
| $L_2$ | 782.4    | 1019.3   | 1200.1   | 100.1         | 99.8          | 100.0         | 9.4           | 7.0           | 5.9           | 95.4°                    | 204.7°                   |
| $L_3$ | 782.8    | 1019.5   | 1199.8   | 99.5          | 99.7          | 100.0         | 11.4          | 13.3          | 8.4           | 219.2°                   | 81.6°                    |
| $L_4$ | 782.7    | 1019.5   | 1200.0   | 99.8          | 100.0         | 100.1         | 11.2          | 12.9          | 8.1           | 210.2°                   | 64.5°                    |

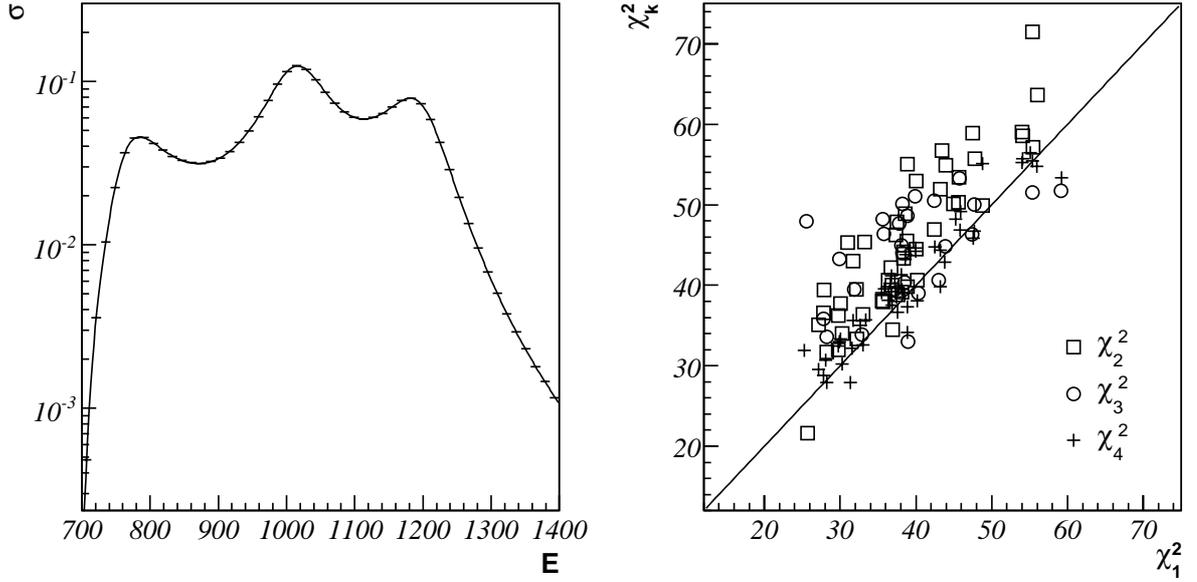

Figure 5. Left: the fit of "experimental" data for cross section (42) with parameters (46). Right: the values $\chi_n^2$ of function $\chi^2$ in local maxima $L_n$ ($n = 2,3,4$) versus the value $\chi_1^2$ in local maximum $L_1$, for 50 numerical experiments with statistics of $10^7$ events.

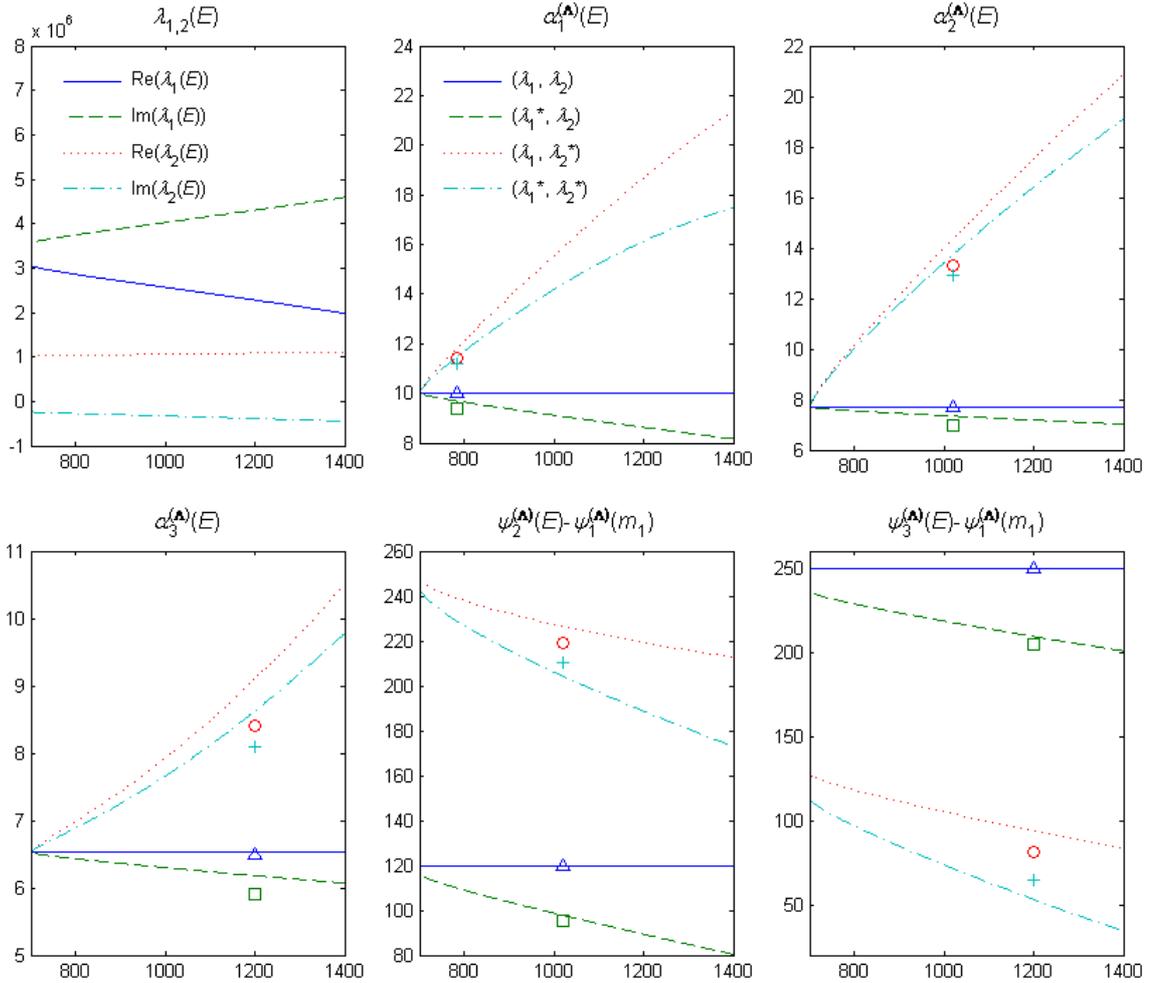

Figure 6. "Roots" $\lambda_1(E)$, $\lambda_2(E)$ and functions $\widetilde{A}_{kx}^{(\Lambda)}(E)$ for the cross section (42) with parameters (46). The markers show average values of parameters in local maxima of the likelihood function: $\triangle - L_1$, $\circ - L_2$, $\square - L_3$, $+ - L_4$.

## 4. Conclusion

The general form of solutions for the parameters of an arbitrary number of interfering Breit-Wigner resonances is found. It is shown that the number of solutions giving the same energy dependence of the cross section, is determined by the positions of the roots of the corresponding characteristic equation. If all roots and complex conjugate numbers are pairwise distinct, then the number of solutions, including the trivial one, is equal to $2^P$, where $P$ is the number of roots. Otherwise, the number of solutions is reduced.

If resonances do not have the Breit-Wigner form, then for any number of resonances, provided that their amplitudes satisfy the certain conditions, the non-trivial solutions exist. However, even in the case when there are no non-trivial solutions, but the shape of resonances weakly differs from the Breit-Wigner one, in the fit of experimental data additional solutions for the parameters, giving the close cross section and corresponding to the solutions for interfering Breit-Wagner resonances, will be found. In case of insufficient statistics, these additional solutions may be indistinguishable by $\chi^2$ from the true one, that is degenerate. The conditions of removal of degeneracy in an experiment are studied in detail in Ref. [5]. Statistics required for removal of degeneracy increase with a decrease of the resonance overlap, as well as with a decrease of the fit range. As seen above, one can also add that the properties of solutions are determined by the properties of the functions $\lambda_k(E)$.

The author is grateful to S.I. Eidelman for useful discussions and help in paper preparation.